\pdfoutput=1
\documentclass[pre,twocolumn,superscriptaddress, showpacs,amsmath,amssymb]{revtex4}
\usepackage{graphicx}
\usepackage{subfigure}
\begin{document}
\title{Dielectric Sensing with Back-Gated Nanowires}
\author{George Boldeiu}
\affiliation{National Institute for Research and Development in Microtechnologies-IMT,
126A, Erou Iancu Nicolae Street, 077190, Bucharest, ROMANIA}
\author{Victor Moagar-Poladian}
\affiliation{National Institute for Research and Development in Microtechnologies-IMT,
126A, Erou Iancu Nicolae Street, 077190, Bucharest, ROMANIA}
\author{Titus Sandu}
\affiliation{National Institute for Research and Development in Microtechnologies-IMT,
126A, Erou Iancu Nicolae Street, 077190, Bucharest, ROMANIA}
\email{titus.sandu@imt.ro}
\date{\today}
\begin{abstract}
Extensive numerical calculations show that the 
capacitance of back-gated nanowires with various degrees of dielectric 
embeddings is accurately described with an effective dielectric constant as 
long as the difference between the dielectric thickness and the 
gate-nanowire distance is held constant. This is valid for dielectrics with 
permittivities ranging from simple air to water. However, due to screening 
the scaling is not valid if the dielectric lies down well below the 
nanowire. Moreover, when only the dielectric thickness varies the 
capacitance characteristics are S-shaped with three distinct regions, of 
which only the first two can be used for dielectric sensing. The first 
region is almost linear while the middle region, with a span of two 
diameters around the center of the nanowire, is the most 
sensitive.
\end{abstract}
\maketitle 
\section{Introduction}

The continuous progress encountered in the interdisciplinary field of 
nanotechnology has made possible the fabrication, characterization, and 
utilization of a plethora of nanosized structures in 1, 2, and 3 dimensions. 
One of the most studied categories of nanostructures is that of nanowires 
(NWs) and, in particular, semiconductor NWs, in which the charged carriers 
are confined in 2 dimensions and can eventually move freely in the third 
direction. The semiconductor NWs offer complex building blocks for devices 
and structures with functionalities used in fields like nanoelectronics or 
photovoltaics \cite{Lieber2011}. Their specific shape makes them ideally as active 
electronic interfaces with biological cells; hence the active structures 
with a NW field effect transistor (FET) are of great promise not only as 
electronic devices but also as interfaces and sensors in biological systems 
\cite{Duan2013}. In addition, electrical measurements and probing of biological cells 
and tissues with the help of NW-FET display high signal-to-noise ratios 
mostly because of protrusion capabilities and small active areas of those 
devices \cite{Timko2010}. 

A metallic-like behavior in NWs is obtained with a doping as large as 
10$^{18}$--10$^{19}$ carriers/cm$^{3}$ where the Debye screening length is 
of a few nanometers (i. e. much smaller than the NW diameter) \cite{Sze1981}. In this 
case the metallic behavior of the NW brings the NW-FET in the linear regime 
where there is a linear relationship between the conductance and the gate 
voltage. In the linear regime there is a definite relation between the 
back-gate capacitance $C_{G}$ of the NW-FET and the effective mobility $\mu 
_{FE} $ which is one of the key parameters characterizing the charge 
transport in the FETs. Thus if we assume that the NW-FET is of length 
$L_{G}$, under a drain-source voltage $V_{DS}$, and a gate-source voltage 
$V_{GS}$ the source-drain current $I_{DS}$ is given by the following 
expression \cite{Dayeh2010}

\begin{equation}
\label{eq1a}
I_{DS} = {\mu _{FE} C_G ({V_{GS} - V_t })V_{DS} }/{L_G^2 },
\end{equation}

\noindent
where $V_{t}$ is the threshold voltage that is defined as the gate voltage 
that fully depletes the free carriers in the NW. The effective mobility $\mu 
_{FE} $ can be calculated from the slope of the current versus gate voltage 
or the transconductance $g_m = {\partial I_{DS} }/{\partial V_{GS} }$ with the following 
formula 

\begin{equation}
\label{eq2a}
\mu _{FE} = \frac{g_m L_G^2 }{C_G V_{DS} }
\end{equation}

Equation (\ref{eq2a}) is used for estimating the charge carrier mobility in NWs and 
shows the key role of the back-gate capacitance. Most of the estimations of 
the back-gate capacitance are based on the assumption that the NW is a long 
cylinder at the distance $t$ above a metallic plane associated with the back 
gate. The capacitance of such a back-gated system is given by

\begin{equation}
\label{eq3a}
C_G = \frac{2\pi \varepsilon _0 \varepsilon _r L_G }{\cosh ^{ - 1}( {t/ R})} 
\end{equation}

\noindent
where $R$ is the radius of the cylinder, and $\varepsilon _r $ is the relative 
dielectric constant of the dielectric in which the system is supposed to be 
totally embedded. 

\begin{figure}[htp]
\includegraphics[width=3.5in,height=1.4in]{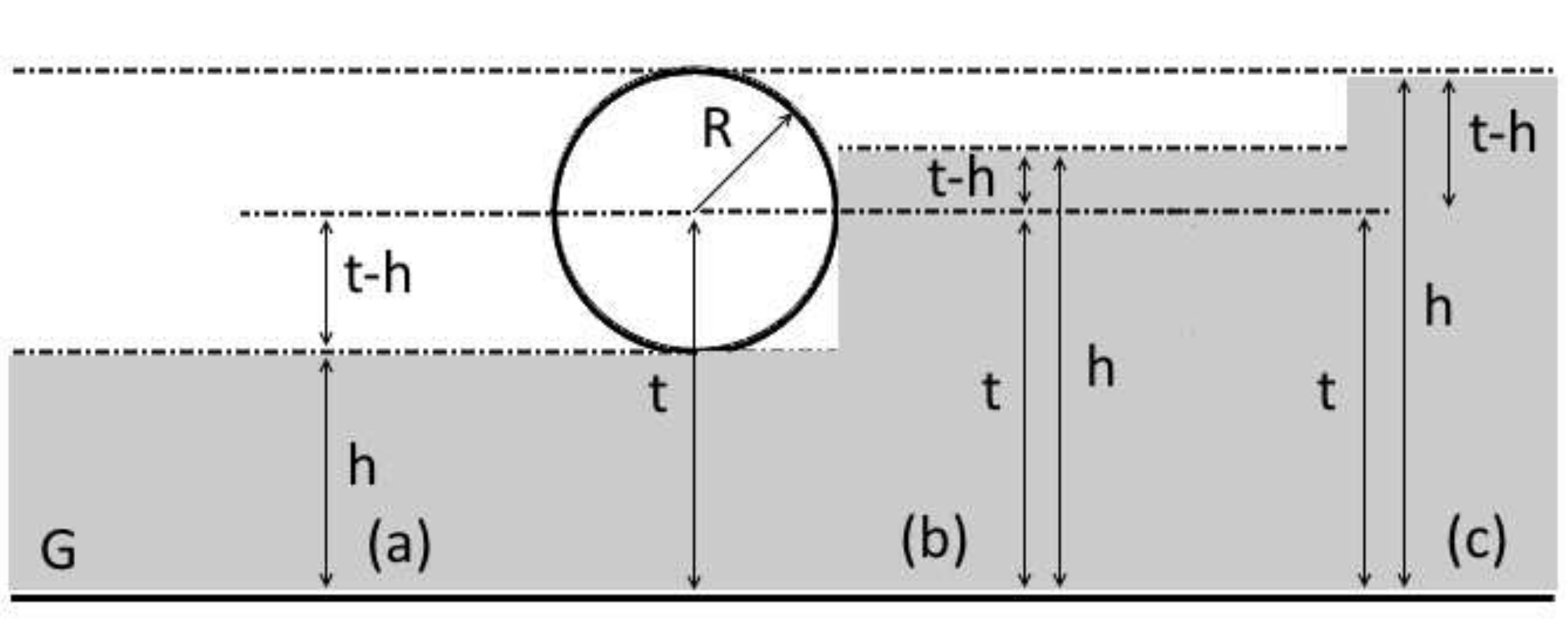}
\caption{Cross-section view of the NW. The NW is depicted as a circle of radius $R$, G is the gate, $t$ is 
the distance from the NW center to the gate and $h$ is the thickness of the dielectric represented as dotted lines.}
  \label{fig:1}
\end{figure}

Typically, the configuration of a NW-FET is shown in Fig. \ref{fig:1}(a), where the NW 
is separated from the gate $G$ by a dielectric slab of thickness $h=t-R$. The 
capacitance formula given by Eq. (\ref{eq3a}) has been commonly used in the 
evaluation of back-gate capacitance of NW-FETs \cite{Wang2003,Duan2001,Javey2002}. On the other hand, Eq. 
(\ref{eq3a}) is correct if only the entire space is a dielectric of permittivity 
$\varepsilon _r $. As a result it was noticed in several works that, for 
real devices, formula given by Eq. (\ref{eq3a}) is valid if the relative permittivity 
of the dielectric slab $\varepsilon _r $ is replaced by an effective 
permittivity $\varepsilon _{eff} $ that is approximately half (more 
precisely 58{\%}) of $\varepsilon _r $ when the dielectric is silicone 
dioxide \cite{Wunnicke2006,Ford2009}. 

In the present work we perform an extensive study by considering not only 
different dielectric slab thicknesses or embeddings but also different 
dielectric constants. Some of the present results have been previously 
reported in a recent conference paper \cite{Boldeiu2014}. Two of those embeddings are 
depicted in Fig. \ref{fig:1}(b) (the NW is more than half-buried by the dielectric) and 
Fig.  \ref{fig:1}(c) with the NW fully buried just below the dielectric-air interface. We 
will show that Eq. (\ref{eq3a}) may be used to describe the back-gate capacitance of 
the NW for arbitrary slab embeddings. This description of the capacitance is 
determined by $t-h$ that is the difference between the distance NW-gate and the 
dielectric thickness. Thus, if $t-h$ is constant, Eq. (\ref{eq3a}) is valid with an 
appropriate $\varepsilon _{eff} $ that depends not only on $t-h$ but also on 
$\varepsilon _r $. Deviation from this behavior occurs when the dielectric 
slab is thinner, i. e., $(t - h)/R \gg 1$. 
We further analyze capacitance changing with respect to dielectric thickness 
in order to assess the sensing capabilities of liquid levels in microfluidic 
systems. We have found that with respect to the dielectric thickness the 
capacitance characteristics have an S-like shape with three distinct 
regions. The capacitance varies almost linearly in the first region, slowly 
in the third one, and rapidly in the middle region that is located around 
the NW. 

The paper has the following structure. In the second section we present the 
problem and the method of resolution. The third section is dedicated to the 
main results and the last section will summarize the conclusions.

\section{Preliminaries and the Method}

Our problem resides in solving the Laplace equation with Dirichlet boundary 
conditions, i. e., a fixed potential $V$ on the surface of cylinder and with a 
grounded back-gate. The capacitance can be calculated numerically by a 
variety of methods including finite element methods \cite{Johnson1987} or boundary element 
methods \cite{Poljak2005}. The boundary element method is the finite element version of 
the boundary integral method which comes from potential theory \cite{Kellog1967}. In the 
static limit metallic regions are considered to totally screen the electric 
fields thus those regions are equipotential regions. This is not valid when, 
instead of static fields, the electromagnetic fields are considered, such 
that metals have a finite dielectric permittivity at optical frequencies. 
Thus, for metallic nanostructures with features much smaller than the 
wavelength of the incoming electromagnetic wave the response of a metallic 
system is given by the same Laplace equation but with different boundary 
conditions (basically there are Neumann boundary conditions) \cite{Sandu2011,Sandu2013}. The 
method is a spectral approach to the boundary integral equation that turns 
out to obtain simultaneously also the electrostatic capacitance of metallic 
nanostructures \cite{Sandu2013b}. 

\begin{figure}[htp]
\includegraphics[width=2.7in,height=1.8in]{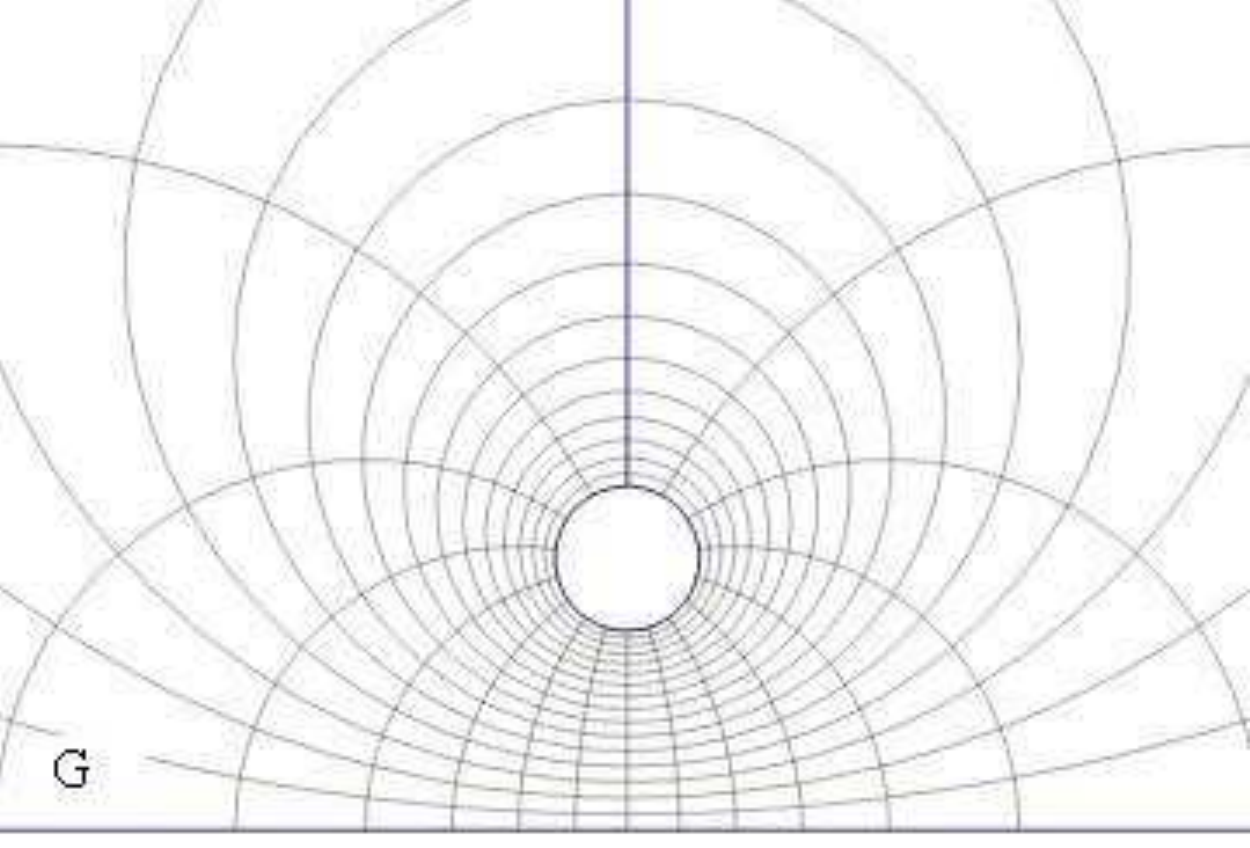}
\caption{The bipolar coordinates ($\eta $, $\xi )$. Surfaces $\eta $ = constant surround the NW and $\xi $ = constant start up on NW and end up on the back-gate G.}
  \label{fig:2}
\end{figure}

Capacitance and the electrostatic response of finite cylinders have been 
studied in  \cite{Sandu2013b} and in  \cite{Sandu2014}, respectively, where explicit expressions for 
both capacitance and electric polarizability of finite cylinders have been 
found. On the other hand, an infinite long cylinder with a grounded planar 
back-gate and homogeneously embedded in a dielectric of relative 
permittivity $\varepsilon _r$ has also an explicit capacitance that is 
given by Eq. (\ref{eq3a}). One calculation procedure is based on the bipolar 
coordinates \cite{Boldeiu2014,Arfken1970}. In the usual setup with the NW perpendicular on the 
(x,y) plane, the bipolar coordinates given by

\begin{equation}
\label{eq4a}
\begin{array}{l}
 x = \frac{a\sinh \left( \eta \right)}{\cosh \left( \eta \right) - \cos 
\left( \xi \right)} \\ 
 y = \frac{a\sin \left( \xi \right)}{\cosh \left( \eta \right) - \cos \left( 
\xi \right)} \\ 
 z = z. \\ 
 \end{array}
\end{equation}

\noindent
generate the equipotential surfaces and the field lines of a cylinder 
charged under a potential $V$ with respect to the grounded planar gate \cite{Boldeiu2014}. 
The equation of the NW surface is $\eta =\eta_{0}$, while the equation of the planar gate 
is $\eta $ = 0. We notice that the ratio $R/t$ from Fig. \ref{fig:1} is ${t/R} = \cosh({\eta 
_0 })$ In the (x,y) plane the new coordinates ($\eta $, $\xi 
)$ are shown in Fig. \ref{fig:2}. Thus, the surfaces $\eta $= constant are cylinders that 
surround our NW \cite{Arfken1970}. They turn out to be equipotential surfaces of a 
charged cylinder including also the surface $\eta $ = 0 that is our planar gate. The 
surfaces $\xi $ = constant are also cylinders and define the field lines that 
start up normally on the NW and end up also normally on the planar gate \cite{Arfken1970}.

We have shown that in the case of electrostatic capacitance once the 
equipotential surfaces and field lines are known one can estimate 
straightforward the capacitance of the system \cite{Sandu2013b}. Accordingly, the 
capacitance of the back-gated NW is

\begin{equation}
\label{eq5a}
C = \varepsilon _0 \varepsilon _r \left( {\int\limits_0^{\eta _0 } 
{\frac{d\eta }{\int\limits_{\eta = constant} {\frac{h_\xi h_z d\xi 
dz}{h_\eta }} }} } \right)^{ - 1}
\end{equation}

\noindent
where $h_{\eta }$, $ h_{\xi }$, and $ h_{z}$ are the Lam\'{e} coefficients of the 
transformation (\ref{eq4a})
\begin{equation}
\label{eq6a}
\begin{array}{l}
 h_\eta = \frac{a}{\cosh \left( \eta \right) - \cos \left( \xi \right)} \\ 
 h_\xi = \frac{a}{\cosh \left( \eta \right) - \cos \left( \xi \right)} \\ 
 h_z = 1. \\ 
 \end{array}
\end{equation}
Combining (\ref{eq5a}) with (\ref{eq6a}) will lead to 

\begin{equation}
\label{eq7a}
C = \frac{2\pi \varepsilon _0 \varepsilon _r L_G }{\eta _0 }
\end{equation}
\noindent
which is just Eq. (\ref{eq3a}) since $\eta _0 = \cosh ^{ - 1}({t/R})$. Similar formulae can be obtained for two non-concentric cylinders 
or for two parallel cylinders \cite{Boldeiu2014}. 

\begin{figure}[htp]
  \begin{center}
   \subfigure {\label{fig:3a} \includegraphics [width=2.8in,height=2.1in]{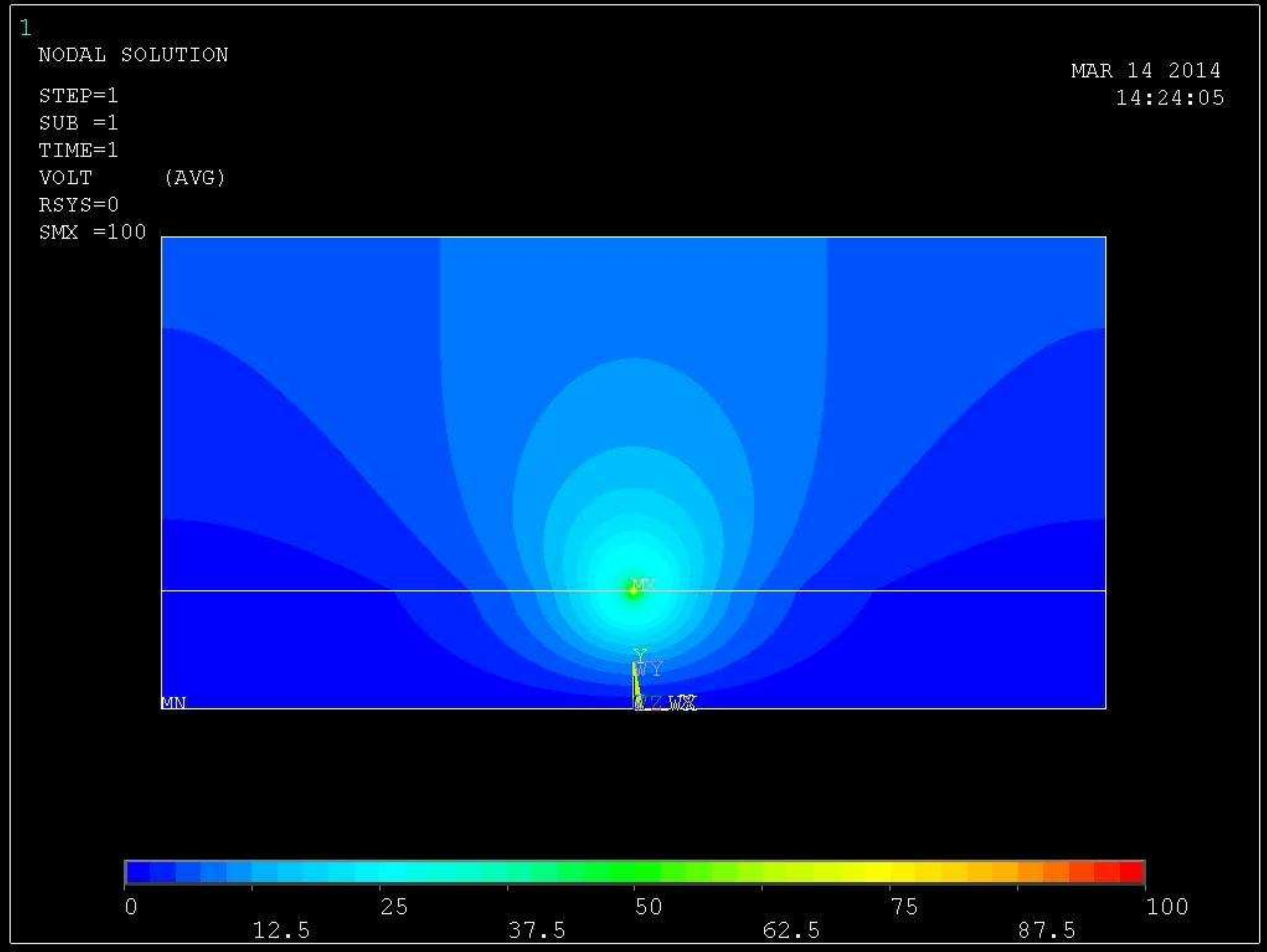}}  \\
    \subfigure {\label{fig:3b} \includegraphics [width=2.80in,height=2.1in]{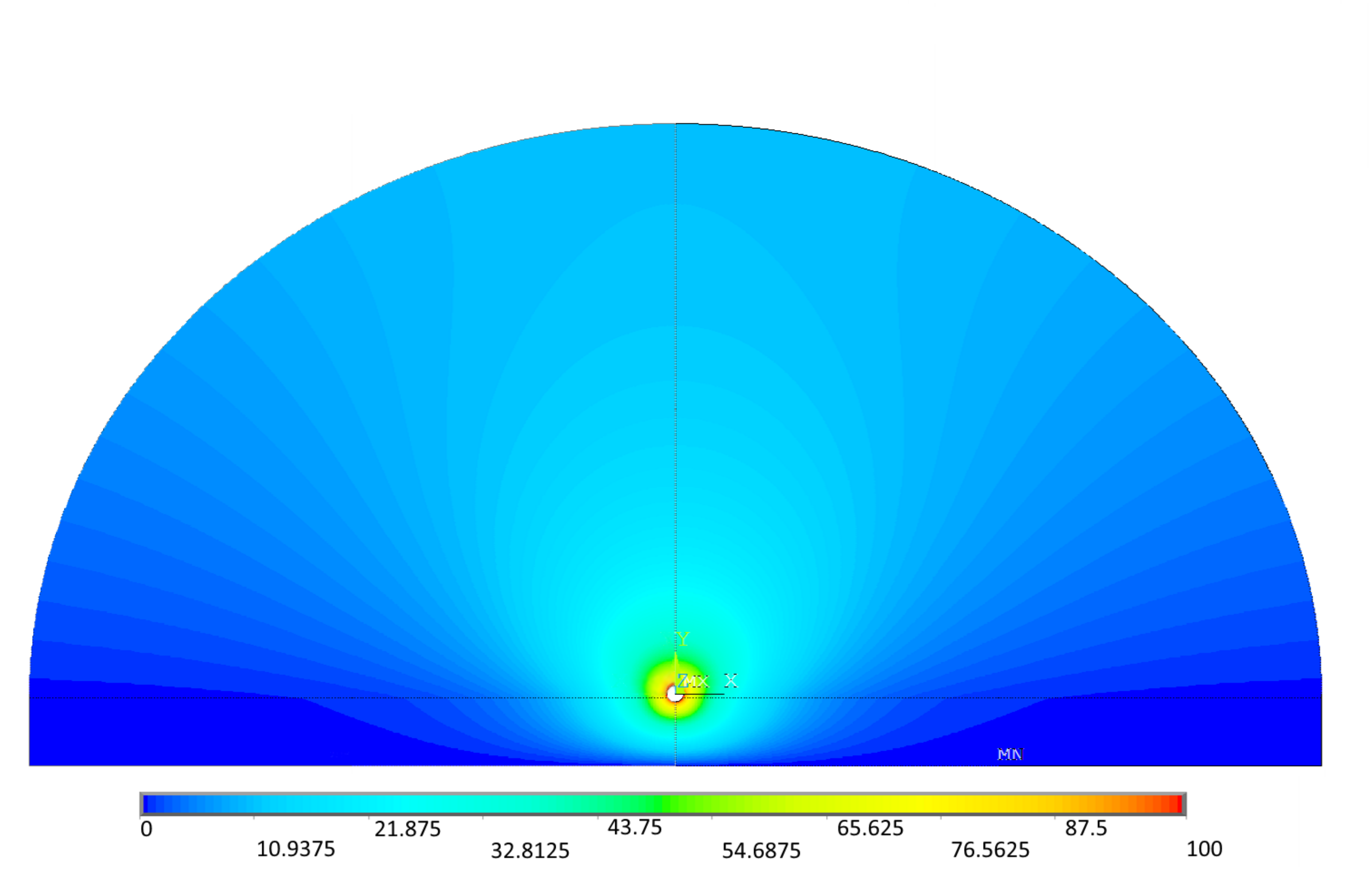}} 
  \end{center}
  \caption{2D cross-section of equipotential surfaces calculated (a) with a finite box and (b) with infinite boundary elements. The dielectric is shown by the horizontal line intersecting the NW which is placed in the middle of the pictures.}
  \label{fig:3}
\end{figure}

Slight modifications of the shapes of nanostructures induce minute changes 
on capacitance \cite{Sandu2013b} or on higher excitations modes \cite{Sandu2012,Sandu2014b}. On the other 
hand, the back-gated NWs are not infinitely long and have a certain degree 
of doping, effects that cannot be analytically quantified. Extensive 
numerical calculations have shown that with a doping as much as 5x10$^{18}$ 
cm$^{ - 3}$ and with an aspect ratio $L_G/R$ as large as 100 the 
behavior of the semiconductor NW is that of infinitely long metallic NW 
\cite{Khanal2007}. Furthermore, the capacitance of the back-gated NW with a finite but 
otherwise arbitrary thickness of the dielectric cannot have a readily 
analytic formula like that of the homogeneous case. Hence we invoke a fully 
numerical procedure to calculate the NW capacitance in various dielectric 
embeddings. Our calculations are based on ANSYS which is a finite element 
based multiphysics software program \cite{ansys}. In ANSYS there are several methods 
for computing the electrostatic capacitance of a metallic system. The first 
and the simplest method is a pure finite element method in a finite 
computation box, hence the equipotential surfaces are enforced to close on 
the boundaries of the computation box (Fig. \ref{fig:3}a). Another method is the 
method that uses infinite boundary elements (Fig. \ref{fig:3}b). As we can see from 
the figure the infinite boundary elements ensure a more physical appearance 
of the equipotential surfaces with respect to a finite computation box by 
considering the same computer overhead. Therefore, the second method is more 
accurate and faster in convergence than the first one. Nevertheless we have 
used the first method but we have tuned it with respect to the second one by 
making the results of the two methods be apart by less than 0.5{\%}. The 
major reason for this choice was the fact that the first method is more 
suitable to be used in ANSYS scripts of repeated calculations.

\begin{figure}[htp]
  \begin{center}
   \subfigure {\label{fig:4a} \includegraphics [width=2.8in,height=2.1in]{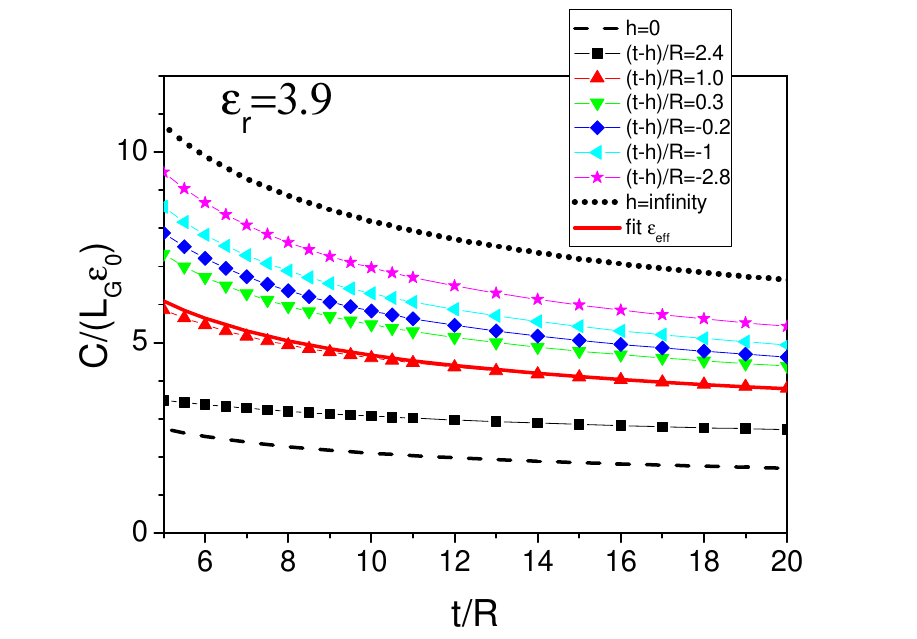}}  \\
    \subfigure {\label{fig:4b} \includegraphics [width=2.80in,height=2.1in]{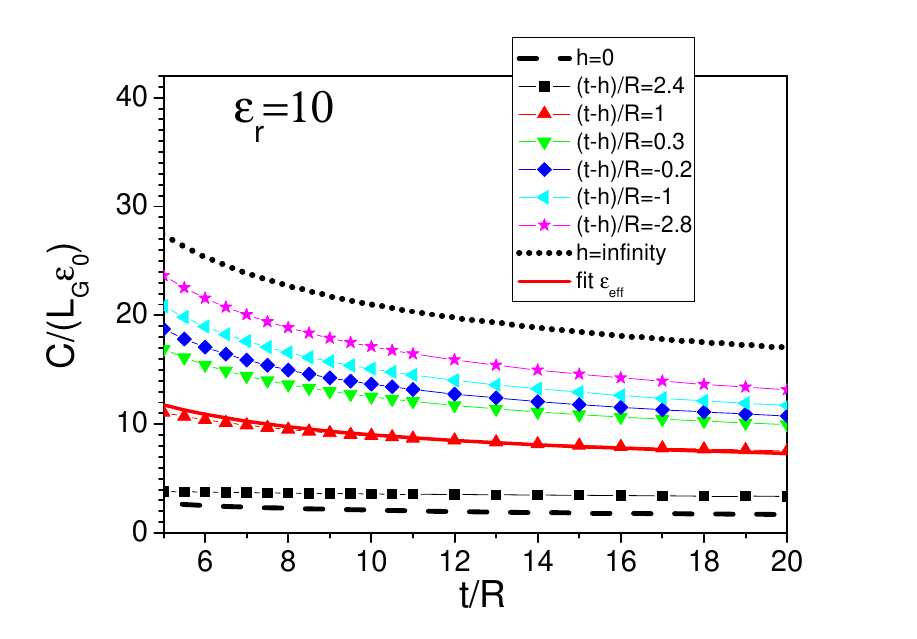}} \\
    \subfigure {\label{fig:4c} \includegraphics [width=2.80in,height=2.1in]{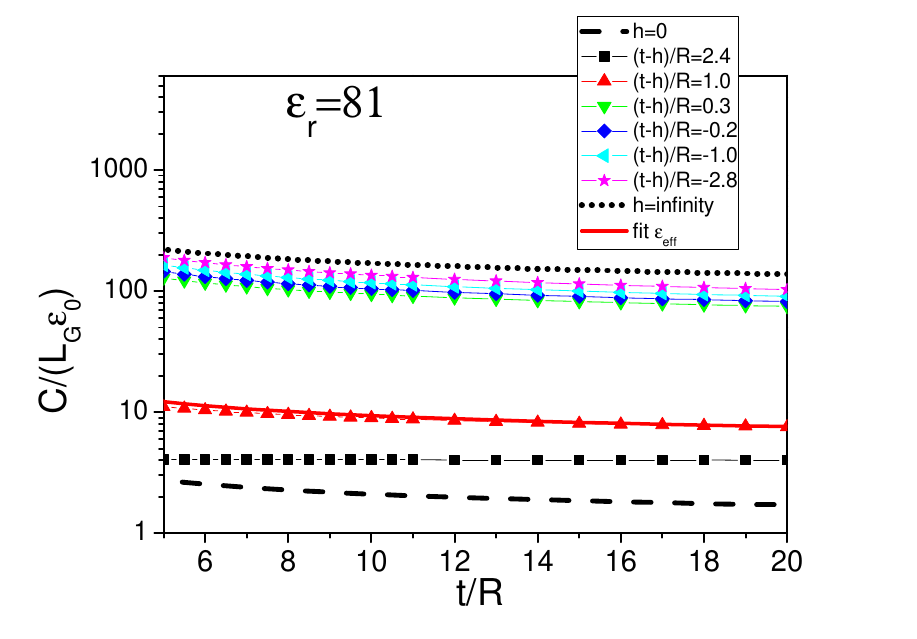}} 
  \end{center}
  \caption{ NW capacitance versus $t/R$ when $t-h$ is kept constant. The curve $h$=0 is the case of air as a dielectric, while the curve $h= $infinity is the case of completely dielectric fillings. The cases (a), (b), and (c) depicted in Fig. \ref{fig:1} are, respectively, $(t-h)/R$ =1, -0.2, -1. The fit of curves for $(t-h)/R$ =1 is plotted with solid lines.}
  \label{fig:4}
\end{figure}

\section{Results}

\begin{figure}[htp]
  \begin{center}
   \subfigure {\label{fig:5a} \includegraphics [width=2.7in,height=2.in]{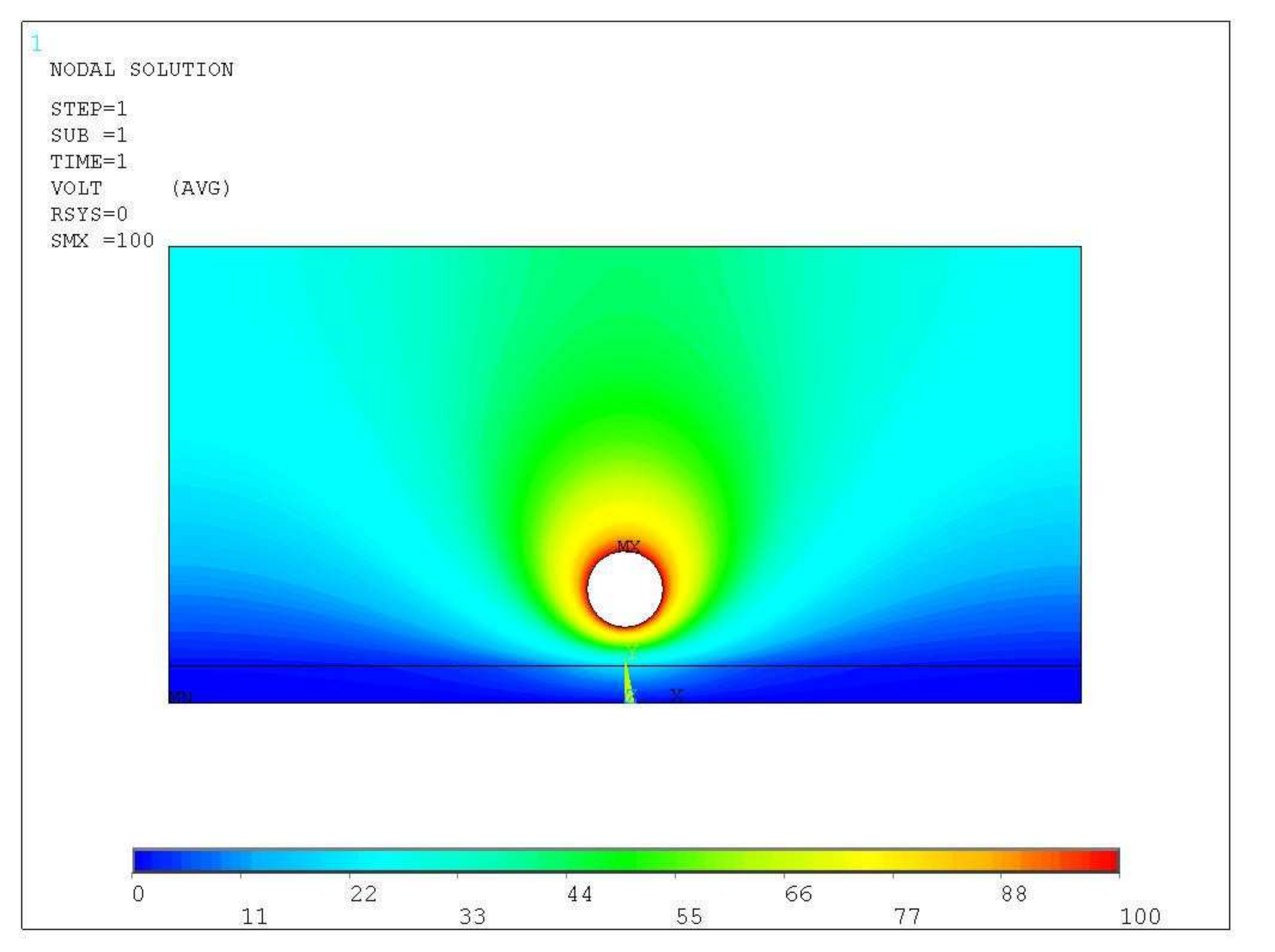}}  \\
    \subfigure {\label{fig:5b} \includegraphics [width=2.7in,height=2.in]{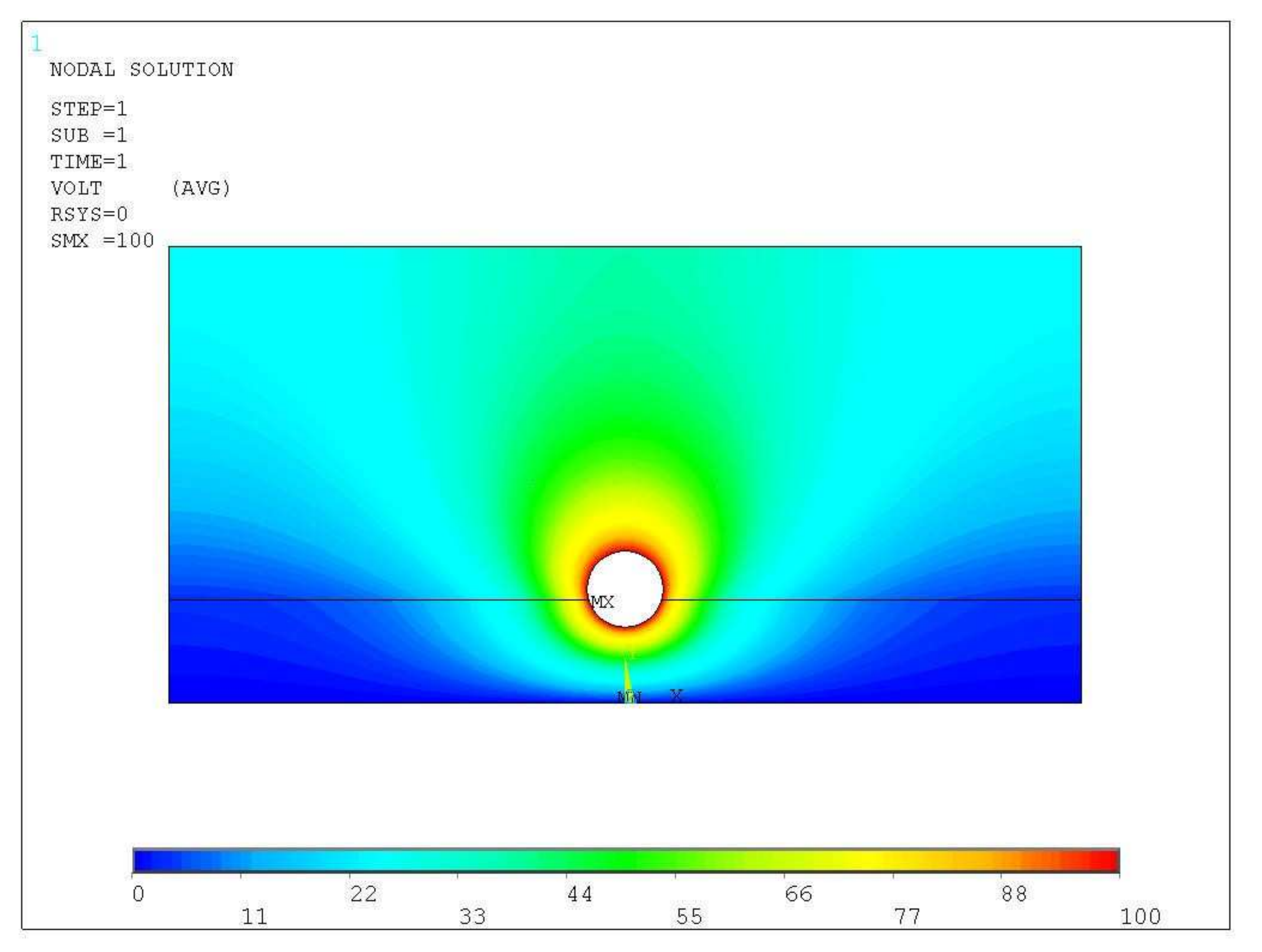}} \\
    \subfigure {\label{fig:5c} \includegraphics [width=2.7in,height=2.in]{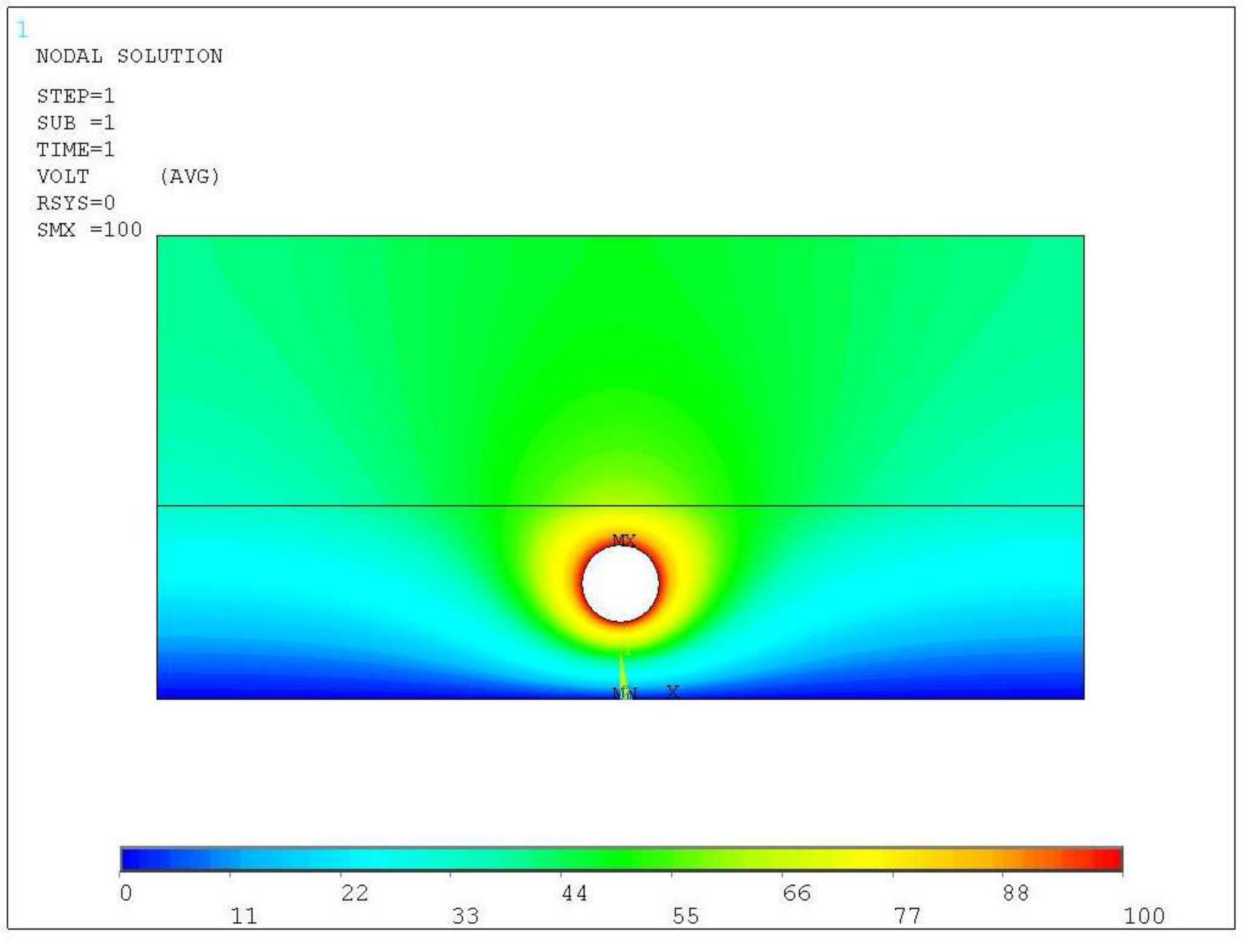}} 
  \end{center}
  \caption{  Cross-section of the equipotential surfaces for dielectric: (a) below the NW, (b) partially burying the NW, and (c) above the NW. Only (a) is able to screen the back gate. We have used $\varepsilon _r = 10$.}
  \label{fig:5}
\end{figure}

\begin{figure}[htp]
  \begin{center}
   \subfigure {\label{fig:6a} \includegraphics [width=2.8in,height=2.1in]{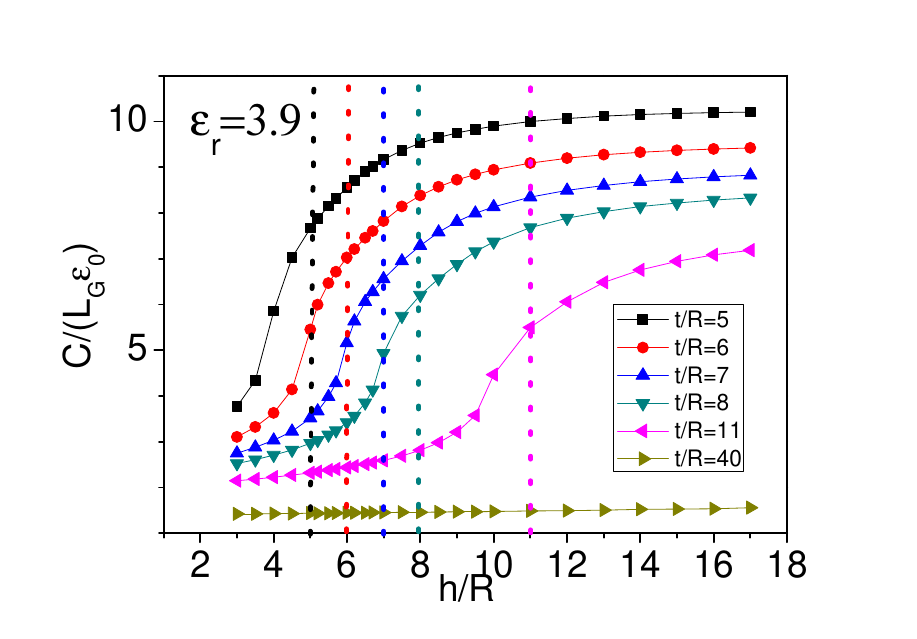}}  \\
    \subfigure {\label{fig:6b} \includegraphics [width=2.80in,height=2.1in]{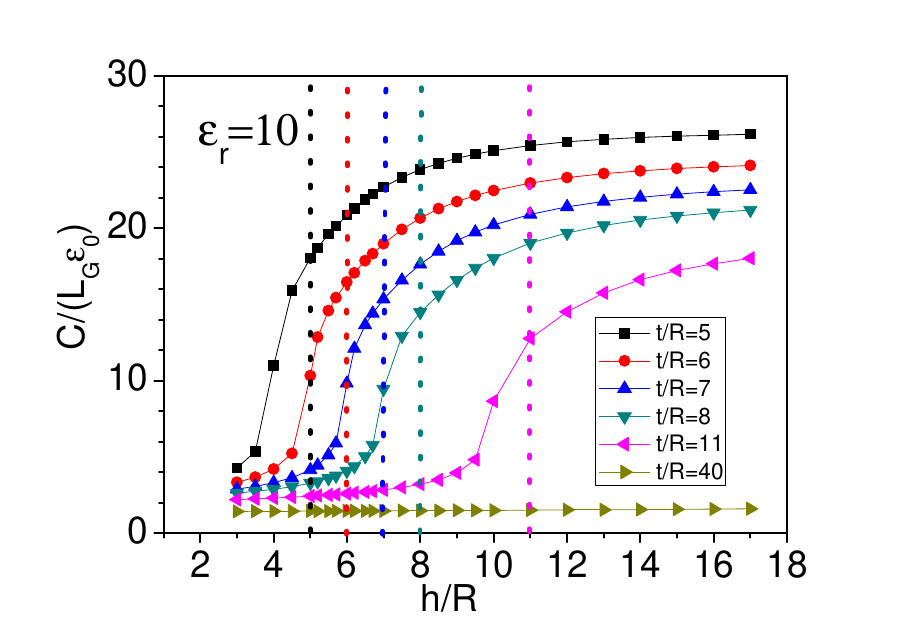}} \\
    \subfigure {\label{fig:6c} \includegraphics [width=2.80in,height=2.1in]{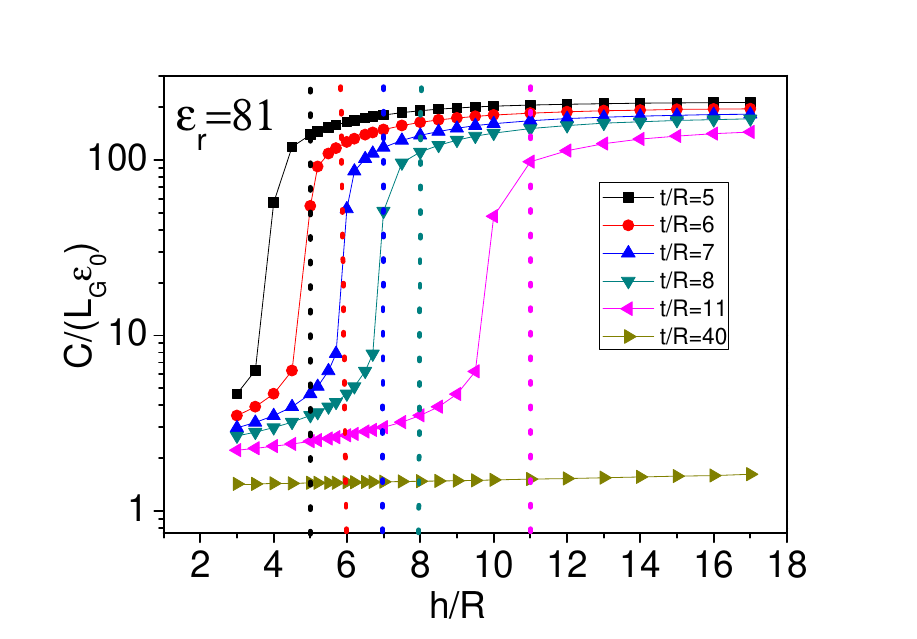}} 
  \end{center}
  \caption{NW capacitance versus h/R when t/R is: 5, 6, 7, 8, 11, and 40. The corresponding vertical lines indicate the ratio t/R in order to delimitate the maximum sensitivity of a NW-FET for dielectric thickness measurements or for proximity sensing.}
  \label{fig:6}
\end{figure}

Our main results are presented in Figs. \ref{fig:4}, \ref{fig:5}, and \ref{fig:6}. The dielectric 
constants considered in this work are $\varepsilon _r = 3.9$ (i. e. 
SiO$_{2})$, $\varepsilon _r = 10$(a lower limit of high-$\kappa $ 
dielectrics), and $\varepsilon _r = 81$ (the dielectric constant of water). 
In Fig. \ref{fig:4} we present the NW capacitance as a function of $t/R$ with $t/R>$5 (the 
typical setting for a NW-FET) and with different dielectric embeddings. The 
level of dielectric embedding was established in such a way that $t-h$ is 
constant. In this setting the capacitance of the NWs shows good scaling 
properties according to Eq. (\ref{eq3a}) with an effective dielectric 
permittivity $\varepsilon _{eff} $ for each value of $t-h$. The particular case 
$(t-h)/R$=1, in which the dielectric separates the NW and the gate plane, is plotted 
with upward oriented triangles and is also fitted with an effective 
dielectric permittivity. The results of fitting are plotted with solid lines 
showing a quite good superposition on the calculated curves. The case 
$(t-h)/R$=1 is the standard configuration of the NW-FET and was studied in \cite{Wunnicke2006} where 
an effective permittivity $\varepsilon _{eff} = 0.57\varepsilon _{SiO_2 } $ 
was given. We have obtained a value $\varepsilon _{eff} = 0.58\varepsilon 
_{SiO_2 } $. In addition, when the high-$\kappa $ dielectric HfO$_{2 }$was 
used ($\varepsilon _r = 25)$ the effective permittivity was about 34{\%} of 
that of the dielectric \cite{Wunnicke2006}. Our calculations indicate that the effective 
permittivities are 43 percent and 5.5 percent of permittivities of the 
dielectric with $\varepsilon _r = 10$ and $\varepsilon _r = 81$, 
respectively. One can further notice that for $(t-h)/R$ varying between -1 and 1 the 
$\varepsilon _{eff} $ varies nonlinearly with respect to the variation of 
$\varepsilon _r $. 

The scaling is still valid for dielectric permittivities as large as that of 
water in spite of the fact that at quite large dielectric constant the 
dielectrics should behave as a metal. Nevertheless the metallic behavior 
(manifested as strong screening) of the dielectric sets in when $(t - 
h)/R \gg 1 $ (i. e., the NW is well above the dielectric 
layer) and the scaling is no longer valid. One can also notice that at 
$(t - h)/R = 2.4$ for both $\varepsilon _r = 10$ and 
$\varepsilon _r = 81$ the capacitance is almost constant, a feature of 
metallic behavior. Fig. \ref{fig:5} illustrates more clearly that only for $(t - h)/R > 1$ (Fig. \ref{fig:5}a) 
the dielectric screens the back gate hence the 
scaling described above cannot longer be valid. In addition, for dielectric 
embeddings that completely cover the NW the capacitance goes very slowly to 
the value of complete filling of the space with dielectric (Fig. \ref{fig:5}c). 

We will further analyze and discuss the previous scaling behavior for 
certain applications used to determine the dielectric thickness by 
electrical measurements using NW-FET. A potential application one can think 
of is the estimation of the liquid height in microfluidic systems. To keep 
the discussion as simple as possible we suppose that the surface of the NW 
is passivated and the liquid is totally wetting the NW surface. In Fig. \ref{fig:6} we 
plotted the NW capacitance with respect to thickness $h$ at various values of 
$t/R$. In fact the curves plotted in Fig. \ref{fig:6} give information about $\varepsilon 
_{eff} $ discussed previously. The capacitance is $S$-shaped with three distinct 
regions. For small thicknesses the capacitance varies linearly as it was 
expected on perturbative grounds. The size of this region increases but the 
slope of the curves decrease with the increase of $t/R$. Hence, at $t/R = 40$ 
the size of this linear region is significantly larger as one can see it 
from Fig. \ref{fig:6}. In this region, in order to exploit its linearity for sensing, 
one has to trade off between the size of the linear region and the 
slope of capacitance with respect to dielectric thickness $h$. By inspecting 
Fig. \ref{fig:6} it seems that $t/R = 10 - 11$ ensures an optimum sensing in this 
region.
The second region displays the largest capacitance variation with greater 
slopes for greater dielectric permittivities. The region spans approximately 
four NW radiuses around the NW center and can be used as a proximity sensor 
in which the capacitance increases considerably when the liquid levels are 
in the vicinity of the NW. Finally, the third region of deeper dielectric 
embedding the capacitance varies smoothly and saturates to its asymptotic 
value of total embedding. 

\section{Conclusions}

In the present work we have studied the back-gate capacitance variation of a 
nanowire with respect to various levels of dielectric embedding. The 
standard configuration in the field effect transistor setting is that in 
which the dielectric separates the nanowire from the gate plane. It was 
previously shown that this standard geometric arrangement has a scaling 
behavior straightforwardly connected to the back-gated nanowire embedded in 
a homogeneous dielectric. The scaling is made with an effective dielectric 
constant that strongly depends on the permittivity of the dielectric. In our 
paper we have shown numerically that this scaling is valid for various 
degrees of dielectric embedding granted the fact that the difference between 
the dielectric thickness and the gate-nanowire distance is constant. The 
scaling is dependent on the difference between the dielectric thickness and 
the gate-nanowire distance and on the permittivity of the dielectric. 
However, the scaling is not valid for dielectric thicknesses much smaller 
than the distance nanowire-gate because of the dielectric screening. We 
further discuss this property for sensing purposes. Thus, we analyze the 
capacitance change with respect to dielectric thickness as an indicator of 
liquid height in microfluidic systems. The capacitance curves have an S-like 
shape with three regions. The first region is almost linear and might be 
optimum for sensing if the ratio between gate-nanowire distance and the 
nanowire radius is about 10. The second region which is of the size of two 
diameters around the center of the nanowire has the largest slope and can be 
used for sensors of proximity. The third region is rather unimportant since 
the capacitance in the region saturates slowly to its asymptotic value given 
by the embedding of the nanowire in a dielectric of infinite thickness.

\begin{acknowledgments}
This work was supported by a grant of the Romanian National Authority for 
Scientific Research, CNCS -- UEFISCDI, project number PNII-ID-PCCE-2011 
-2-0069. 
\end{acknowledgments}


\begin{thebibliography}{23}
\expandafter\ifx\csname natexlab\endcsname\relax\def\natexlab#1{#1}\fi
\expandafter\ifx\csname bibnamefont\endcsname\relax
  \def\bibnamefont#1{#1}\fi
\expandafter\ifx\csname bibfnamefont\endcsname\relax
  \def\bibfnamefont#1{#1}\fi
\expandafter\ifx\csname citenamefont\endcsname\relax
  \def\citenamefont#1{#1}\fi
\expandafter\ifx\csname url\endcsname\relax
  \def\url#1{\texttt{#1}}\fi
\expandafter\ifx\csname urlprefix\endcsname\relax\def\urlprefix{URL }\fi
\providecommand{\bibinfo}[2]{#2}
\providecommand{\eprint}[2][]{\url{#2}}

\bibitem[{\citenamefont{Lieber}(2011)}]{Lieber2011}
\bibinfo{author}{\bibfnamefont{C.~M.} \bibnamefont{Lieber}},
  \bibinfo{journal}{MRS Bull.} \textbf{\bibinfo{volume}{36}},
  \bibinfo{pages}{1052} (\bibinfo{year}{2011}).

\bibitem[{\citenamefont{Duan et~al.}(2013)\citenamefont{Duan, Fu, Liu, and
  Lieber}}]{Duan2013}
\bibinfo{author}{\bibfnamefont{X.}~\bibnamefont{Duan}},
  \bibinfo{author}{\bibfnamefont{T.~M.} \bibnamefont{Fu}},
  \bibinfo{author}{\bibfnamefont{J.}~\bibnamefont{Liu}}, \bibnamefont{and}
  \bibinfo{author}{\bibfnamefont{C.~M.} \bibnamefont{Lieber}},
  \bibinfo{journal}{Nano Today} \textbf{\bibinfo{volume}{8}},
  \bibinfo{pages}{351} (\bibinfo{year}{2013}).

\bibitem[{\citenamefont{Timko et~al.}(2010)\citenamefont{Timko, Cohen-Karni,
  Qing, Tian, and Lieber}}]{Timko2010}
\bibinfo{author}{\bibfnamefont{B.~P.} \bibnamefont{Timko}},
  \bibinfo{author}{\bibfnamefont{T.}~\bibnamefont{Cohen-Karni}},
  \bibinfo{author}{\bibfnamefont{Q.}~\bibnamefont{Qing}},
  \bibinfo{author}{\bibfnamefont{B.}~\bibnamefont{Tian}}, \bibnamefont{and}
  \bibinfo{author}{\bibfnamefont{C.~M.} \bibnamefont{Lieber}},
  \bibinfo{journal}{IEEE Trans. Nanotechnology} \textbf{\bibinfo{volume}{9}},
  \bibinfo{pages}{269} (\bibinfo{year}{2010}).

\bibitem[{\citenamefont{Sze}(1981)}]{Sze1981}
\bibinfo{author}{\bibfnamefont{S.~M.} \bibnamefont{Sze}},
  \emph{\bibinfo{title}{Physics of Semiconductor Devices}}
  (\bibinfo{publisher}{Wiley Inter-Science}, \bibinfo{address}{New York},
  \bibinfo{year}{1981}), \bibinfo{edition}{2nd} ed.

\bibitem[{\citenamefont{Dayeh}(2010)}]{Dayeh2010}
\bibinfo{author}{\bibfnamefont{S.~A.} \bibnamefont{Dayeh}},
  \bibinfo{journal}{Semicond. Sci. Technol.} \textbf{\bibinfo{volume}{25}},
  \bibinfo{pages}{024004} (\bibinfo{year}{2010}).

\bibitem[{\citenamefont{Wang et~al.}(2003)\citenamefont{Wang, Wang, Javey, Tu,
  Dai, McIntyre, Krishnamohan, and Saraswat}}]{Wang2003}
\bibinfo{author}{\bibfnamefont{D.}~\bibnamefont{Wang}},
  \bibinfo{author}{\bibfnamefont{Q.}~\bibnamefont{Wang}},
  \bibinfo{author}{\bibfnamefont{A.}~\bibnamefont{Javey}},
  \bibinfo{author}{\bibfnamefont{R.}~\bibnamefont{Tu}},
  \bibinfo{author}{\bibfnamefont{H.}~\bibnamefont{Dai}},
  \bibinfo{author}{\bibfnamefont{H.~K. P.~C.} \bibnamefont{McIntyre}},
  \bibinfo{author}{\bibfnamefont{T.}~\bibnamefont{Krishnamohan}},
  \bibnamefont{and} \bibinfo{author}{\bibfnamefont{K.~C.}
  \bibnamefont{Saraswat}}, \bibinfo{journal}{Appl. Phys. Lett.}
  \textbf{\bibinfo{volume}{83}}, \bibinfo{pages}{2432} (\bibinfo{year}{2003}).

\bibitem[{\citenamefont{Duan et~al.}(2001)\citenamefont{Duan, Cui, Wang, and
  Lieber}}]{Duan2001}
\bibinfo{author}{\bibfnamefont{X.}~\bibnamefont{Duan}},
  \bibinfo{author}{\bibfnamefont{Y.~H.~Y.} \bibnamefont{Cui}},
  \bibinfo{author}{\bibfnamefont{J.}~\bibnamefont{Wang}}, \bibnamefont{and}
  \bibinfo{author}{\bibfnamefont{C.~M.} \bibnamefont{Lieber}},
  \bibinfo{journal}{Nature(London)} \textbf{\bibinfo{volume}{409}},
  \bibinfo{pages}{66} (\bibinfo{year}{2001}).

\bibitem[{\citenamefont{Javey et~al.}(2002)\citenamefont{Javey, Kim, Brink,
  Wang, Ural, Guo, McIntyre, McEuen, Lundstrom, and Dai}}]{Javey2002}
\bibinfo{author}{\bibfnamefont{A.}~\bibnamefont{Javey}},
  \bibinfo{author}{\bibfnamefont{H.}~\bibnamefont{Kim}},
  \bibinfo{author}{\bibfnamefont{M.}~\bibnamefont{Brink}},
  \bibinfo{author}{\bibfnamefont{Q.}~\bibnamefont{Wang}},
  \bibinfo{author}{\bibfnamefont{A.}~\bibnamefont{Ural}},
  \bibinfo{author}{\bibfnamefont{J.}~\bibnamefont{Guo}},
  \bibinfo{author}{\bibfnamefont{P.}~\bibnamefont{McIntyre}},
  \bibinfo{author}{\bibfnamefont{P.}~\bibnamefont{McEuen}},
  \bibinfo{author}{\bibfnamefont{M.}~\bibnamefont{Lundstrom}},
  \bibnamefont{and} \bibinfo{author}{\bibfnamefont{H.}~\bibnamefont{Dai}},
  \bibinfo{journal}{Nature Materials} \textbf{\bibinfo{volume}{1}},
  \bibinfo{pages}{241} (\bibinfo{year}{2002}).

\bibitem[{\citenamefont{Wunnicke}(2006)}]{Wunnicke2006}
\bibinfo{author}{\bibfnamefont{O.}~\bibnamefont{Wunnicke}},
  \bibinfo{journal}{Appl. Phys. Lett.} \textbf{\bibinfo{volume}{89}},
  \bibinfo{pages}{083102} (\bibinfo{year}{2006}).

\bibitem[{\citenamefont{Ford et~al.}(2009)\citenamefont{Ford, Ho, Chueh, Tseng,
  Fan, Guo, Bokor, and Javey}}]{Ford2009}
\bibinfo{author}{\bibfnamefont{A.~C.} \bibnamefont{Ford}},
  \bibinfo{author}{\bibfnamefont{J.~C.} \bibnamefont{Ho}},
  \bibinfo{author}{\bibfnamefont{Y.~L.} \bibnamefont{Chueh}},
  \bibinfo{author}{\bibfnamefont{Y.~C.} \bibnamefont{Tseng}},
  \bibinfo{author}{\bibfnamefont{Z.}~\bibnamefont{Fan}},
  \bibinfo{author}{\bibfnamefont{J.}~\bibnamefont{Guo}},
  \bibinfo{author}{\bibfnamefont{J.}~\bibnamefont{Bokor}}, \bibnamefont{and}
  \bibinfo{author}{\bibfnamefont{A.}~\bibnamefont{Javey}},
  \bibinfo{journal}{Nano Lett.} \textbf{\bibinfo{volume}{9}},
  \bibinfo{pages}{360} (\bibinfo{year}{2009}).

\bibitem[{\citenamefont{Boldeiu et~al.}(2014)\citenamefont{Boldeiu,
  Moagar-Poladian, and Sandu}}]{Boldeiu2014}
\bibinfo{author}{\bibfnamefont{G.}~\bibnamefont{Boldeiu}},
  \bibinfo{author}{\bibfnamefont{V.}~\bibnamefont{Moagar-Poladian}},
  \bibnamefont{and} \bibinfo{author}{\bibfnamefont{T.}~\bibnamefont{Sandu}}, in
  \emph{\bibinfo{booktitle}{IEEE-International Semiconductor Conference (CAS)}}
  (\bibinfo{address}{Sinaia, Romania}, \bibinfo{year}{2014}), pp.
  \bibinfo{pages}{273--276}.

\bibitem[{\citenamefont{Johnson}(1987)}]{Johnson1987}
\bibinfo{author}{\bibfnamefont{C.}~\bibnamefont{Johnson}},
  \emph{\bibinfo{title}{Numerical Solutions of Partial Differential Equations
  by Finite Element Method}} (\bibinfo{publisher}{Cambridge, University Press},
  \bibinfo{address}{Cambridge, UK}, \bibinfo{year}{1987}).

\bibitem[{\citenamefont{Poljak and Brebbia}(2005)}]{Poljak2005}
\bibinfo{author}{\bibfnamefont{D.}~\bibnamefont{Poljak}} \bibnamefont{and}
  \bibinfo{author}{\bibfnamefont{C.~A.} \bibnamefont{Brebbia}},
  \emph{\bibinfo{title}{Boundary Element Methods for Electrical Engineers}}
  (\bibinfo{publisher}{WIT}, \bibinfo{address}{Boston, USA},
  \bibinfo{year}{2005}).

\bibitem[{\citenamefont{Kellog}(1967)}]{Kellog1967}
\bibinfo{author}{\bibfnamefont{O.~D.} \bibnamefont{Kellog}},
  \emph{\bibinfo{title}{Foundations of Potential Theory}}
  (\bibinfo{publisher}{Springer-Verlag}, \bibinfo{address}{Berlin, Heidelberg,
  New York}, \bibinfo{year}{1967}).

\bibitem[{\citenamefont{Sandu et~al.}(2011)\citenamefont{Sandu, Vrinceanu, and
  Gheorghiu}}]{Sandu2011}
\bibinfo{author}{\bibfnamefont{T.}~\bibnamefont{Sandu}},
  \bibinfo{author}{\bibfnamefont{D.}~\bibnamefont{Vrinceanu}},
  \bibnamefont{and}
  \bibinfo{author}{\bibfnamefont{E.}~\bibnamefont{Gheorghiu}},
  \bibinfo{journal}{Plasmonics} \textbf{\bibinfo{volume}{6}},
  \bibinfo{pages}{407} (\bibinfo{year}{2011}).

\bibitem[{\citenamefont{Sandu}(2013)}]{Sandu2013}
\bibinfo{author}{\bibfnamefont{T.}~\bibnamefont{Sandu}},
  \bibinfo{journal}{Plasmonics} \textbf{\bibinfo{volume}{8}},
  \bibinfo{pages}{391} (\bibinfo{year}{2013}).

\bibitem[{\citenamefont{Sandu et~al.}(2013)\citenamefont{Sandu, Boldeiu, and
  Moagar-Poladian}}]{Sandu2013b}
\bibinfo{author}{\bibfnamefont{T.}~\bibnamefont{Sandu}},
  \bibinfo{author}{\bibfnamefont{G.}~\bibnamefont{Boldeiu}}, \bibnamefont{and}
  \bibinfo{author}{\bibfnamefont{V.}~\bibnamefont{Moagar-Poladian}},
  \bibinfo{journal}{J. Appl. Phys} \textbf{\bibinfo{volume}{114}},
  \bibinfo{pages}{224904} (\bibinfo{year}{2013}).

\bibitem[{\citenamefont{Sandu}(2014)}]{Sandu2014}
\bibinfo{author}{\bibfnamefont{T.}~\bibnamefont{Sandu}},
  \bibinfo{journal}{Proceedings of the Romanian Academy, Series A}
  \textbf{\bibinfo{volume}{15}}, \bibinfo{pages}{338} (\bibinfo{year}{2014}).

\bibitem[{\citenamefont{Arfken}(1970)}]{Arfken1970}
\bibinfo{author}{\bibfnamefont{G.}~\bibnamefont{Arfken}},
  \emph{\bibinfo{title}{Mathematical Methods for Physicists}}
  (\bibinfo{publisher}{Academic Press}, \bibinfo{address}{Orlando, Florida},
  \bibinfo{year}{1970}), \bibinfo{edition}{2nd} ed.

\bibitem[{\citenamefont{Sandu}(2012)}]{Sandu2012}
\bibinfo{author}{\bibfnamefont{T.}~\bibnamefont{Sandu}},
  \bibinfo{journal}{Journal of Nanoparticle Research}
  \textbf{\bibinfo{volume}{14}}, \bibinfo{pages}{905} (\bibinfo{year}{2012}).

\bibitem[{\citenamefont{Sandu and Boldeiu}(2014)}]{Sandu2014b}
\bibinfo{author}{\bibfnamefont{T.}~\bibnamefont{Sandu}} \bibnamefont{and}
  \bibinfo{author}{\bibfnamefont{G.}~\bibnamefont{Boldeiu}},
  \bibinfo{journal}{Digest Journal of Nanomaterials and Biostructures}
  \textbf{\bibinfo{volume}{9}}, \bibinfo{pages}{1255} (\bibinfo{year}{2014}).

\bibitem[{\citenamefont{Khanal and Wu}(2007)}]{Khanal2007}
\bibinfo{author}{\bibfnamefont{D.~R.} \bibnamefont{Khanal}} \bibnamefont{and}
  \bibinfo{author}{\bibfnamefont{J.}~\bibnamefont{Wu}}, \bibinfo{journal}{Nano
  Lett.} \textbf{\bibinfo{volume}{7}}, \bibinfo{pages}{2778}
  (\bibinfo{year}{2007}).



\bibitem[{ans()}]{ansys}
\bibinfo{note}{ANSYS® Academic Research, Release 12.1, Help System, Low-
  Frequency Electromagnetic Guide, ANSYS, Inc.}

\end{thebibliography}

\end{document}